\documentclass[aps,showpacs,preprintnumbers,amsmath,amssymb,superscriptaddress,floatfix,a4paper,nofootinbib,11pt]{revtex4}
\usepackage[a4paper,left=22mm,right=22mm,top=25mm,bottom=25mm]{geometry}
\usepackage{lipsum}
\usepackage{graphicx}
\usepackage{epsfig}
\usepackage{epstopdf}
\usepackage{hyperref}
\usepackage{amsmath}
\usepackage{amsfonts}
\usepackage{amssymb}
\usepackage{amsfonts,color}

\begin{document}

\title{Unification of inflation and dark matter in the Higgs-Starobinsky model}
\author{Daris Samart}
\email{dsamart82@gmail.com/daris.s@chula.ac.th}
\affiliation{High Energy Physics Theory Group, Department of Physics, Faculty of Science, Chulalongkorn University, Phyathai Rd., Bangkok 10330, Thailand}

\author{Phongpichit Channuie}
\email{channuie@gmail.com}
\affiliation{College of Graduate Studies \& School of Science, Walailak University, Thasala, Nakhon Si Thammarat, 80160, Thailand}

\date{\today}

\vskip 1pc
\begin{abstract}
In this work, we study unified picture of inflation and dark matter in the Higgs-Starobinsky (HS) model. As pointed out in the literature, Starobinsky $R^2$ inflation is induced by quantum correction effect from the large Higgs-curvature (graviton) coupling. We start with non-minimal coupling HS action in Jordan frame. We then transform the Jordan frame action into the Einstein one using the conformal transformation. The inflation potential is derived from the gravitational action of non-minimal-Higgs coupling and Starobinsky term in Einstein frame where the $R^2$ term is dominated in the inflationary phase of the universe. For model of inflation, we compute the inflationary parameters and confront them with Planck 2015 data. We discover that the predictions of the model are in excellent agreement with the Planck analysis. In addition, we study the Higgs field as a candidate for dark matter. The renormalization group equations (RGEs) of Higgs-Starobinsky scenario with standard model at one-loop is qualitatively analyzed. By using the solutions of parameter spaces from RGE analysis, the remnants of the quantum effect ($R^2$ term) from Higgs-graviton coupling will be implied by dark matter relic abundance.
\end{abstract}

\date{today}

\maketitle

\section{introduction}
An inflationary scenario is a well-established paradigm describing an early universe and posts an indispensable ingredient of modern cosmology. Regarding degrees of freedom contained in the SM of particle physics or quantum general relativity, Higgs inflation and Starobinsky model have received much attraction in the recent years. The preceding scenario technically requires a large non-minimal coupling $(\xi)$ between a Higgs boson $H$ and the Ricci scalar $R$, i.e. $\xi H^{\dagger}H R$ which leads to successful inflation and produces the spectrum of primordial fluctuations in good agreement with the observational data. The formulation of the later model is based on $R^{2}$ gravity. It is worth noting that these two models are minimalistic and nicely compatible with the latest Planck data \cite{Ade:2015lrj}.  

As pointed out in Ref.\cite{Calmet:2016fsr}, both operators $\xi H^{\dagger}H R$ and $R^{2}$ are expected to be generated when the SM of particle physics is coupled to general relativity.  More importantly, due to a large non-minimal coupling of the Higgs boson and the Ricci scalar, Starobinky inflation can be generated by quantum effects. In this situation, the Higgs boson need not to start at a high field value at inflation. In addition, in the HS model, a Higgs potential can be stabilized.

However, the true nature of inflation is still unclear. Apart form inflation, the nature of dark matter conveys one
of the unsolved problems in physics and also dark energy is still the greatest cosmic mystery. There was a large number of models proposed so far possible to account for DM candidates \cite{Feng:2010gw,Feng:2003uy,Viel:2005qj} and references therein. The aim of this work is to present a unified description of inflation and dark matter in the context of the Higgs-Starobinsky model. Regrading the present investigation, the feature is multi-fold:
\begin{itemize}
\item (i) There is no need of physics beyond the SM of particle physics since the operators here are expected to be generated when general relativity is coupled to the SM of particle physics.

\item (ii) We propose a cosmological scenario that unifies comic inflation and dark matter to a single framework. The model is minimalistic.

\item (ii) Regarding this two-field scenario, the model dose not suffer from the unitarity problem as that of the Higgs inflation.

\end{itemize}

This paper is organized as follows: In Sec.\ref{sec2}, we take a short recap of the Higgs-Starobinsky model. We also consider quantum corrections and renormalized group equations. In Sec.\ref{sec3}, we study inflationary implications of the HS model. Here we first construct inflationary model, compute its inflationary observables and then compare with the latest Planck data. In Sec.\ref{sec4}, we present a model of dark matter and make qualitative discussion of the model. Discussions and conclusions are given in the last section.

\section{Model Set-up}
\label{sec2}
\subsection{Inducing Starobinsky, $R^2$ term by non-minimal Higgs coupling}
In this section, the Higgs-Starobinsky action will be considered and constructed. As mentioned earlier, the Higgs-Starobinsky mechanisms is induced by quantum effect with the large coupling of Higgs-curvature \cite{Calmet:2016fsr,Salvio:2015kka}. we start with the gravitational action of the Higgs-curvature coupling with self-interacting Higgs field, it reads
\begin{eqnarray}
S_J = \int d^4x\sqrt{-g}\,\left[ \frac{1}{2}\,M_P^2\, R + \frac{1}{2}\,\xi\,\sigma^2\,R
+ \frac{1}{2}\,g^{\mu\nu}\partial_\mu\sigma\,\partial_\nu\sigma - \frac{\lambda}{4}\sigma^4  \right],
\label{H-J}
\end{eqnarray}
where the subscript $S_J$ stands for the action in Jordan frame and $M_P = 1/\sqrt{8\pi G}$ and $\xi$ are Planck mass and non-minimal Higgs coupling constant, respectively. The $\sigma$ field is scalar field with the conventional Higgs potential and its self-interacting coupling strength $\lambda$. In the framework of quantum field theory in curved spacetime, it is well known that one needs to introduce the pure higher gravitational terms into the action for making proper renormalization procedure \cite{Birrell:1982ix,Parker:2009uva}. Introducing the higher gravitational correction terms, the gravitational action in this work is written by \cite{Markkanen:2018bfx,Ghilencea:2018rqg},
\begin{eqnarray}
S_J &=& \int d^4x\sqrt{-g}\,\left[ - \frac{1}{2}\,M_P^2\, R - \frac{1}{2}\,\xi\,\sigma^2\,R
+ \frac{1}{2}\,g^{\mu\nu}\partial_\mu\sigma\,\partial_\nu\sigma - \frac{\lambda}{4}\sigma^4  \right]
\nonumber\\
&+& \int d^4x\sqrt{-g}\,\left[ -c_1\,R^2 - c_2\,R_{\mu\nu}\,R^{\mu\nu} - c_3\,R_{\mu\nu\rho\sigma}R^{\mu\nu\rho\sigma}\right]
\label{HOD1-J}\\
&=& \int d^4x\sqrt{-g}\,\left[ -\frac{1}{2}\,M_P^2\, R - \frac{1}{2}\,\xi\,\sigma^2\,R
+ \frac{1}{2}\,g^{\mu\nu}\partial_\mu\sigma\,\partial_\nu\sigma - \frac{\lambda}{4}\sigma^4  - \alpha\,R^2 - \beta\,\mathcal{C}^2 - \gamma\,\mathcal{G}^2\right],
\label{HOD2-J}
\end{eqnarray}
where the $\mathcal{C}^2$ and $\mathcal{G}^2$ terms are Wely tensor and Guass-Bonnet term, respectively and they are defined by
\begin{eqnarray}
\mathcal{C}^2 &=& \frac13\,R^2 - 2\,R_{\mu\nu}\,R^{\mu\nu} + R_{\mu\nu\rho\sigma}R^{\mu\nu\rho\sigma} \,,
\nonumber\\
\mathcal{G}^2 &=& R^2 - 4\,R_{\mu\nu}\,R^{\mu\nu} + R_{\mu\nu\rho\sigma}R^{\mu\nu\rho\sigma} \,.
\end{eqnarray}
In addition, the couplings, $\alpha$\,, $\beta$ and $\gamma$ are linear combinations of the $c_{1,2,3}$ as
\begin{eqnarray}
\alpha = c_1 + \frac13\left( c_2 + c_3 \right)   \,,\qquad \beta = \frac12\,c_2 + 2\,c_3  \,\qquad \gamma =  -\frac12\,c_2 - c_3\,.
\end{eqnarray}
We note that the parameters  $\alpha$, $\beta$ and $\gamma$ vanish at the classical level. As mentioned, the higher order curvature terms of these parameters are introduced for cancelling the divergence in the energy-momentum tensor when the perturbative expansions of the loop diagrams are taken into account \cite{Birrell:1982ix,Parker:2009uva}. Since we work in a flat FLRW universe, it is worth noting that the Weyl tensor vanishes in the flat FLRW metric and the Guass-Bonnet term is topological invariant in four-dimensional spacetime which is also vanished. Only the relevant higher order curvature term of the action in this work is $R^2$\,. Then, the gravitational action of Higgs-Starobinsky (HS) model in Jordan frame reads,
\begin{eqnarray}
S_J = \int d^4x\sqrt{-g}\,\left[ -\frac{1}{2}\,M_P^2\, R - \frac{1}{2}\,\xi\,\sigma^2\,R - \alpha \,R^2
+ \frac{1}{2}\,g^{\mu\nu}\partial_\mu\sigma\,\partial_\nu\sigma - \frac{\lambda}{4}\sigma^4  \right].
\label{HS-J}
\end{eqnarray}
Moreover, the $R^2$ will contribute to improve the effective potential at one-loop level and we will discuss about its quantum effect and RGEs latter. For more detail about quantum effects of the $R^2$ term in the curved spacetime, we recommend the readers to Refs. \cite{Birrell:1982ix,Parker:2009uva,Markkanen:2018bfx,Ghilencea:2018rqg}. In this work, the Higgs-Starobinsky mechanisms is introduced in Refs.\cite{Calmet:2016fsr,Salvio:2015kka} where the non-minimal Higgs-curvature coupling is a trigger of inflation with large value of the non-minimal coupling constant. This mechanism generates the Starobinsky inflation. With the Higgs-Starobinsky model, one does not need to set the Higgs field values in the early universe to be large. Then, according to the observed values of the standard model Higgs vacuum, we achieve the corresponding stability of the electroweark vacuum with the experiments \cite{Calmet:2016fsr}. This model dose not suffer from the unitarity problem of the Higgs inflation also. In this work, we will study the inflation by using this picture. The Higgs-Starobinsky model has been studied in Refs.\cite{Bezrukov:2011gp,Kaneda:2015jma,Ruf:2017xon,Salvio:2017oyf,Barbon:2015fla,Wang:2017fuy,Ema:2017rqn,Pi:2017gih,He:2018gyf,Gorbunov:2018llf,
Gundhi:2018wyz,Karam:2018mft,Enckell:2018uic} and most of the works in the literatures focused in the multi-field inflation framework. With the small value of the Higgs field and in the slow-roll regime during inflation, i.e., suppressing kinetic term of the Higgs field, one can integrate out the non-minimal Higgs coupling term, $\xi\,\sigma^2\,R$\, via the equation of motion which gives $\sigma_c^2 = -6\,\xi\,R/\lambda$\,. This has been shown in Ref.\cite{He:2018gyf} that the action in Eq.(\ref{HS-J}) becomes,
\begin{eqnarray}
S_J &=& \int d^4x\sqrt{-g}\,\left[ -\frac{1}{2}\,M_P^2\, R - \frac{M_P^2}{12\,M^2} \,R^2  \right],
\label{Star-J}
\\
M^2 &=& \frac{M_P^2}{12\left(\alpha + 3\,\xi^2/(2\,\lambda)\right)}\,.
\end{eqnarray}
The above action is usual Starobinsky inflation action. In the Einstein frame, one defines the scalaron mass of the Starobinsky inflaton field as
\begin{eqnarray}
M_\alpha^2 = \frac{M_P^2}{12\,\alpha}\,,
\end{eqnarray} and the scalaron (inflation) mass is modified by the following relation \cite{He:2018gyf}
\begin{eqnarray}
M^2 = \frac{M_\alpha^2}{1+18\,(\xi^2/\lambda)\,M_\alpha^2/M_P^2}\,. \label{Mde}
\end{eqnarray}
According to the observational constraints of the amplitudes of the curvature perturbation, one finds $M \approx 1.3\times 10^{-5}M_P$ \cite{Faulkner:2006ub}. By using the fixing $M$ parameter, we obtain the relation between three parameters $\xi$, $\alpha$ and $\lambda$ and we will employ action in Eq.(\ref{Star-J}) to work out relevant inflation parameters and fix the parameters from the Higgs-Starobinsky model with the observational data in the next section.

\subsection{Quantum corrections and renomalization group equations}
In this subsection, we will briefly review and discuss the non-minimal Higgs coupling induced the Starobinsky $R^2$ term. After that we will close this subsection with the renormalization group equations at one-loop of the standard model parameters in the presence of curved spacetime \cite{Parker:2009uva,Markkanen:2018bfx}. The non-minimal Higgs coupling induced $R^2$ has been shown in Refs.\cite{Calmet:2016fsr,Salvio:2015kka}. In this work, we follow the Higgs-Starobinsky mechanism and briefly review that how the $R^2$ term is induced as shown in \cite{Calmet:2016fsr,Ghilencea:2018rqg}.
\subsubsection{Pure gravitational terms and Higgs field}
We start with the action in Eq.(\ref{HOD1-J}). By using the dimensional regularization scheme via heat kernel technique, the one-loop effective potential (with absorbtion of the wave function renormalization constant) of the Eq.(\ref{HOD1-J}) is given by \cite{Ghilencea:2018rqg,Markkanen:2018bfx},
\begin{eqnarray}
V_{\rm eff}(\sigma) &=& -\frac{\xi}{2}\,R\,\sigma^2 + \frac{\lambda}{4}\,\sigma^4 - M_P^2\,R - c_1\,R^2 - c_2\,R_{\mu\nu}\,R^{\mu\nu} - c_3\,R_{\mu\nu\rho\tau}R^{\mu\nu\rho\tau}
\nonumber\\
&+& \frac{1}{64\pi^2}\left( 3\,\lambda\,\sigma^2 -\left( \xi + \frac16\right)R \right)\left[ \log\left( \frac{\left| 3\,\lambda\,\sigma^2 -\left( \xi + \frac16\right)R \right|}{\mu^2}\right) - \frac32 \right]
\nonumber\\
&+& \frac{1}{64\pi^2}\frac{\left( R_{\mu\nu\rho\tau}R^{\mu\nu\rho\tau} - R_{\mu\nu}\,R^{\mu\nu}\right)}{90}\left[ \log\left( \frac{\left| 3\,\lambda\,\sigma^2 -\left( \xi + \frac16\right)R \right|}{\mu^2}\right) \right],
\label{1loop-pot}
\end{eqnarray}
where $\mu$ is renormalization (subtraction) scale. In this subsection, we note that the couplings appear in above potential are represented for running couplings due to the quantum correction effects as,
\begin{eqnarray}
c_{1,2,3} = c_{1,2,3}(\sigma)\,,\qquad \xi = \xi(\sigma)\,,\qquad \lambda = \lambda(\sigma)\,.
\end{eqnarray}
For the classical couplings, they are written by the following forms,
\begin{eqnarray}
c_{1,2,3}^{(0)} = c_{1,2,3}(\sigma=0)\,,\qquad \xi^{(0)} = \xi(\sigma=0)\,,\qquad \lambda^{(0)} = \lambda(\sigma=0)\,.
\end{eqnarray}
We refer Refs.\cite{Ghilencea:2018rqg,Markkanen:2018bfx} for detail calculation of the effective potential in Eq.(\ref{1loop-pot}) and general concepts and techniques in quantum field in curved spacetime see textbooks \cite{Birrell:1982ix,Parker:2009uva}. Applying one-loop effective potential in Eq.(\ref{1loop-pot}) to the Callan-Symanzik equation, one obtains relevant beta functions of the couplings $c_{1,2,3}$ as \cite{Ghilencea:2018rqg,Markkanen:2018bfx},
\begin{eqnarray}
\beta_{c_1} = -\frac{1}{16\pi^2}\frac{1}{2}\left(\xi + \frac16\right)^2,\qquad 
\beta_{c_2} = -\frac{1}{16\pi^2}\left(\frac{1}{180}\right),\qquad
\beta_{c_3} = \frac{1}{16\pi^2}\left(\frac{1}{180}\right)\,,
\label{gravity-beta}
\end{eqnarray}
where the beta function $\beta_F$ is defined by,
\begin{eqnarray}
\beta_F = \frac{dF}{dt} \,,\qquad \quad t = \ln \mu\,.
\end{eqnarray}
The solutions of the beta functions in Eq.(\ref{gravity-beta}) are given by \cite{Ghilencea:2018rqg},
\begin{eqnarray}
c_1 &=& c_1^{(0)} -  \frac{1}{16\pi^2}\left(\frac{1}{2}\right)\left( \xi + \frac16\right)^2\ln\left(\frac{\sigma}{\mu}\right),
\nonumber\\
c_2 &=& c_2^{(0)} - \frac{1}{16\pi^2}\left(\frac{1}{180}\right)\ln\left(\frac{\sigma}{\mu}\right) ,
\nonumber\\
c_3 &=& c_3^{(0)} + \frac{1}{16\pi^2}\left(\frac{1}{180}\right)\ln\left(\frac{\sigma}{\mu}\right) .
\end{eqnarray}
Then, we can re-write the gravitational action in Eq.(\ref{HS-J}) in terms of the running couplings (with quantum corrections at one-loop level) as,
\begin{eqnarray}
S_J = \int d^4x\sqrt{-g}\,\left[ -\frac{1}{2}\,M_P^2\, R - \frac{1}{2}\,\xi(\sigma)\,\sigma^2\,R - \alpha(\sigma) \,R^2
+ \frac{1}{2}\,g^{\mu\nu}\partial_\mu\sigma\,\partial_\nu\sigma - \frac{\lambda(\sigma)}{4}\sigma^4  \right].
\label{HS2-J}
\end{eqnarray}
the $\alpha(\sigma)$ coupling is linear combinations of the $c_{1,2,3}(\sigma)$ and it reads,
\begin{eqnarray}
\alpha(\sigma) = c_1(\sigma) + \frac13\left( c_2(\sigma) + c_3(\sigma) \right)  = \alpha^{(0)} + \frac{1}{16\pi^2}\left( \xi + \frac16\right)^2\ln\left( \sqrt{\frac{\mu}{\sigma}}\,\right).
\label{alpha-sol}
\end{eqnarray}
Next we will consider the $\alpha(\sigma)$ parameter in case of $\alpha^{(0)} = 0$ (at tree level of quantum loop expansion). In the case of setting the renormalization scale at Planck mass $\mu \approx M_P$ and at sub-Planckian field $\sigma \ll M_P$, one finds,
\begin{eqnarray}
\alpha(\sigma) = \frac{1}{16\pi^2}\left( \xi + \frac16\right)^2\ln\left( \sqrt{\frac{\sigma}{\mu}}\,\right)\,.
\label{alpha-xi}
\end{eqnarray}
This result shows that the $\alpha$ coupling of the $R^2$ term is generated by the non-minimal Higgs coupling, $\xi$ at the one-loop level \cite{Calmet:2016fsr,Ghilencea:2018rqg}. At fixed and small Higgs field regime $\sigma_0^2 \ll M_P^2/\xi$\,, on one hand, the $\xi\,\sigma_0^2\,R$ in the action (\ref{HS2-J}) is suppressed. On the other hand, the $\alpha\,R^2$ dominates the action. We finally achieve the usual Starobinsky inflation. As was shown in Eq.(\ref{alpha-xi}), in addition, the $\alpha$ coupling is written in terms of the non-minimal coupling parameter. One can tune or fit the $\xi$ rather than $\alpha$ and we will constrain the value of the $\xi$ with inflation from observational data in the latter.

\subsubsection{One-loop renormalization group equations for standard model}
The Starobinki inflation generated from non-minimal Higgs coupling term has been discussed and demonstrated in the previous subsection. More completely, we will extend our study to the renormalization group equations for the standard model of particle physics in the presence of the curved spacetime. Results in this subsection discussing below will be very useful for the study of dark matter in the Higgs-Starobinsky model. 

Our goal in this subsection is to obtain the running coupling constants of the standard model parameters with the presence of the gravitational couplings, $\alpha$ and $\xi$. We follow the main results from Refs.\cite{Markkanen:2018bfx,Markkanen:2018pdo}. By using heat kernel technique and dimensional regularization scheme, the Callan-Symanzik equation of the effective potential at one-loop for the standard model with the Starobinsky $R^2$ term leads to the renormalization group equations given below \cite{Markkanen:2018bfx}. For gravitational part, the beta-functions are given by \cite{Markkanen:2018bfx}
\begin{eqnarray}
\beta_\xi &=& \frac{1}{16\pi^2}\left(\xi + \frac16\right)\left( 12\,\lambda + 2\,Y_2 - \frac32\left( g'\right)^2 - \frac92\,g^2\right),
\label{SM-xi}\\
\beta_\alpha &=& \frac{1}{16\pi^2}\left(\frac{1}{3}\right)\left(\xi + \frac16\right)^2\,,
\label{SM-alpha}
\end{eqnarray}
We note that the RGE of the $\alpha$ coupling is modified by the presence of standard model with coefficient $\frac13$ instead of $\frac12$ in the pure gravitational part. The beta-functions of the matters and fields are given by \cite{Markkanen:2018bfx}
\begin{eqnarray}
\beta_{y_t} &=& \frac{1}{16\pi^2}\,y_t\left( \frac32\left( y_t^2 - y_b^2\right) + Y_2 -\frac{1}{12}\Big(17(g')^2 + 27 g^2 + 96g_3^3 \Big)\right)
\label{SM1}\\
\beta_{y_b} &=& \frac{1}{16\pi^2}\,y_b\left( \frac32\left( y_b^2 - y_t^2\right) + Y_2 -\frac{1}{12}\Big(5(g')^2 + 27 g^2 + 96g_3^3 \Big)\right)
\label{SM2}\\
\beta_{y_l} &=& \frac{1}{16\pi^2}\,y_l\left( \frac32y_l^2 + Y_2 -\frac{1}{12}\Big(45(g')^2 + 27 g^2 \Big)\right)
\label{SM3}\\
\beta_{\lambda} &=& \frac{1}{16\pi^2}\,\lambda\left( 24\lambda^2 -3\lambda\Big( (g')^2 + 3g^2\Big) +\frac38\Big( (g')^4 + 2(g')^2g^2 + 3g^4\Big) 
+ 4\lambda Y_2 -2 Y_4\right)
\label{SM4}\\
\beta_{m^2} &=& \frac{1}{16\pi^2}\,m^2\left( 12\lambda -\frac32(g')^2 -\frac92\,g^2 + 2 Y_2 \right)
\label{SM5}\\
\beta_{g'} &=& \frac{1}{16\pi^2}\left( \frac{41}{6}\right)(g')^3 \,,\qquad
\beta_{g} = \frac{1}{16\pi^2}\left( -\frac{19}{6}\right)g^3 \,,\qquad
\beta_{g_3} = \frac{1}{16\pi^2}\left( -7\right)g_3^4 \,, 
\label{SM6}
\end{eqnarray} 
where 
\begin{eqnarray}
Y_2 &=& 3\Big( y_u^2 + y_c^2 + y_t^2 \Big) + 3\Big( y_d^2 + y_s^2 + y_b^2 \Big) + \Big( y_e^2 + y_\mu^2 + y_\tau^2 \Big) \,,
\\
Y_4 &=& 3\Big( y_u^4 + y_c^4 + y_t^4 \Big) + 3\Big( y_d^4 + y_s^4 + y_b^4 \Big) + \Big( y_e^4 + y_\mu^4 + y_\tau^4 \Big) \,.
\end{eqnarray}
We use $y_{u,d,s,c,t,b}$ for Yukawa coupling of the $u,\,d,\,s,\,c,\,t,\,b$ quarks and $y_{l=e,\mu,\tau}$ is Yukawa coupling of electron, muon and tau. The $m$ and $\lambda$ are mass and self-interaction coupling of the Higgs. The parameters, $g'$, $g$ and $g_3$ stand for U(1), SU(2) and SU(3) guage couplings. More detail and discussion of the one-loop effective potential and other related quantities, we refer to Refs.\cite{Markkanen:2018bfx,Markkanen:2018pdo}. We will solve relevant beta-functions given in this subsection for using to study the relic abundances of psuedoscalar Higgs sector as candidate of the dark matter with standard model in the latter section.

\section{Inflationary implication from the HS model}
\label{sec3}
In this section, we will study of the inflation in the Higgs-Starobinsky model. In the inflation phase of the universe, we have shown in the previous section that the Starobinsky $R^2$ term dominates the action in Eq.(\ref{HS-J}). Then, we will use the Starobinsky inflation model in this section. Moreover, the Einstein frame is used as the physical frame in this work. By using an usual conformal transformation, the action in Eq.(\ref{Star-J}) is written in the Einstein frame in the following form,
\allowdisplaybreaks
\begin{eqnarray}
S_E = \int d^4x\,\sqrt{-\widetilde g}\,\left[ -\frac12\,M_P^2\widetilde{R} + \frac12\,\widetilde{g}^{\mu\nu}\,\partial_\mu\varphi\,\partial_\nu\varphi - V(\varphi) \right],
\label{action-E}
\end{eqnarray}
where all quantities with ``\,\,$\widetilde{~}$\,\," are represented quantities in Einstein frame. The conformal factor, $\Omega^2$ plays important role on transformation of the gravitational action from Jordan frame to Einstein frame. The relation between metric tensors of Jordan and Einstein frames reads,
\begin{eqnarray}
g_{\mu\nu} = \Omega^2\,\widetilde{g}_{\mu\nu}\,.
\end{eqnarray}
The corresponding conformal factor of Eq.(\ref{HS-J}) is given by
\begin{eqnarray}
\Omega^2 = \frac{2}{M_P^2}\,\frac{\partial}{\partial R}\left( \frac12\,M_P^2\,R + \frac{M_P^2}{12\,M^2}\,R^2\right) = 1 + \frac{R}{3\,M^2} \,,
\end{eqnarray}
where the definition of the effective mass $M$ is given in Eq.(\ref{Mde}). The Ricci scalar in Jordan frame is written in terms of quantities in Einstein frame as
\begin{eqnarray}
R = \Omega^2\left(\widetilde{R} + 3\,\widetilde{g}^{\mu\nu}\partial_\mu\partial_\nu \ln\Omega^2
- \frac{3}{2}\,\widetilde{g}^{\mu\nu}\partial_\mu\ln\Omega^2\,\partial_\nu\ln\Omega^2 \right).
\end{eqnarray}
More importantly, the scalaron field, $\varphi$ of the Starobinsky inflation is introduced via
\begin{eqnarray}
\varphi = M_P\sqrt{\frac32}\,\ln \Omega^2 \,.
\end{eqnarray}
Using the definition of the scalaron field, one can write the effective potential of the scalaron in Einstein frame as
\begin{eqnarray}
V(\varphi) = \frac{3}{4}\,M_P^2\, M^2
\left( 1- e^{-\sqrt{\frac23}\,\frac{\varphi}{M_P}} \right)^2 .
\label{S-pot}
\end{eqnarray}
This is usual Starobinsky scalaron potential in Einstein frame and we will employ this potential in the analysis of inflation below. According to Higgs-Starobinsky mechanism as shown in Eq.(\ref{alpha-sol}), the condition $\varphi<M_P$ is kept throughout this analysis.
\subsection{Slow-roll approximation}
We are ready to study inflation in this subsection. A flat homogeneous and isotropic FLRW metric is used as the background of the universe and it is written by
\begin{eqnarray}
ds^2 = dt^2 - a(t)^2\,\Big( d^2x + d^2y + d^2z \Big)\,, 
\end{eqnarray}  
where $t$ in this subsection is cosmic time and $a(t)$ is the scale factor of the universe. The corresponding Friedman equation, its cosmic time derivative and the Klein-Gordon equation of the scalaron from above metric and the action in Eq.(\ref{action-E}) are
\begin{eqnarray}
H^2 &=& \frac{1}{3\,M_P^2}\left( \frac12\dot{\varphi}^2 + V(\phi)\right),
\\
\dot{H} &=& -\frac{1}{3\,M_P^2}\left( \frac12\dot{\varphi}^2 \right),
\\
\ddot{\varphi} &=& - 3\,H\,\dot{\varphi} - \partial_\varphi\,V(\varphi) \,,
\end{eqnarray}
where $H \equiv \dot a/a$ and $\dot a \equiv da/dt$\,. In the slow-roll regime, the kinetic term of the scalaron (inflaton) is varying very slow with respect to the cosmic time and it is suppressed. The Friedman equation can be re-written as
\begin{eqnarray}
H^2 \approx \frac{1}{3\,M_P^2}V(\phi) = \frac14\,M^2\left( 1 - e^{-\sqrt{\frac23}\,\frac{\varphi}{M_P}}\right)^2 .
\end{eqnarray}
Next we recall the definitions of the slow roll parameters and they read
\begin{eqnarray}
\epsilon &=& \frac{M_P^2}{2} \left( \frac{1}{V(\varphi)}\frac{dV(\varphi)}{d \varphi} \right)^2 , \qquad\qquad
\eta = M_P^2 \left( \frac{1}{V(\varphi)}\frac{d^2V(\varphi)}{d \varphi^2} \right), \qquad
\nonumber\\
\zeta &=& M_P^2 \left( \frac{1}{V(\varphi)}\frac{dV(\varphi)}{d \varphi}\frac{d^3V(\varphi)}{d \varphi^3} \right), \qquad\qquad
N = \frac{1}{M_P^2} \int _{\varphi_{\rm end}} ^{\varphi_N} \frac{V(\varphi)}{dV(\varphi) /d\varphi} d \varphi \,.
\label{slow-roll}
\end{eqnarray}
Applying the scalaron potential in Eq.  to the slow roll parameters (\ref{slow-roll}), we obtain
\begin{eqnarray}
\epsilon = \frac{4}{3\left[e^{\sqrt{\frac23}\,\frac{\varphi}{M_P}} - 1\right]^2}\,,\quad 
\eta = - \frac{4\left[e^{\sqrt{\frac23}\,\frac{\varphi}{M_P}} - 2\right]}{3\left[e^{\sqrt{\frac23}\,\frac{\varphi}{M_P}} - 1\right]^2}\,,
\quad
\zeta =\frac{16\left[e^{\sqrt{\frac23}\,\frac{\varphi}{M_P}} - 4\right]}{9\left[e^{\sqrt{\frac23}\,\frac{\varphi}{M_P}} - 1\right]^3}\,.
\end{eqnarray}
At inflation end i.e., $\epsilon_{\rm end} = 1$\,, we find $\varphi_{\rm end} = 0.764\,M_P$\,.
The number of e-folding number from $\varphi_{\rm end}$ to $\varphi_{N}$ is given by
\begin{eqnarray}
N = \frac34\left[e^{\sqrt{\frac23}\,\frac{\varphi_{N}}{M_P}} - e^{\sqrt{\frac23}\,\frac{\varphi_{\rm end}}{M_P}} -\frac{\sqrt{6}}{4\,M_P}\left( \varphi_N - \varphi_{\rm end}\right) \right]\approx \frac34 e^{\sqrt{\frac23}\,\frac{\varphi_{N}}{M_P}} \,,
\end{eqnarray}
where we have used the field values $\varphi_{N}\gg \varphi_{\rm end}$. We approximately find $\varphi_{N}\sim \sqrt{3/2}M_{P}\ln(4N/3)$. With $N=60$, we obtain $\varphi_N \approx 5.37\,M_P$\,.
\subsection{Contact with observational Constraints}
Next we constrain our scalaron potential with the COBE normalization condition for fixing parameters in the Higgs-Starobinsky model. We use reads,
\begin{eqnarray}
\frac{V(\varphi)}{\epsilon} \simeq (0.0276\,M_P)^4\,.
\end{eqnarray}
Using the potential in (\ref{S-pot}), we get,
\begin{eqnarray}
\xi\simeq 2.3\times 10^{-3} \sqrt{\lambda } \sqrt{1.8475\times 10^{10} N^2-128648. \alpha }
\end{eqnarray}
In order for $\xi$ to satisfy the usual Higgs inflation, i.e. $\xi\sim 10000$, we discover
\begin{eqnarray}
\lambda\sim -\frac{1.5\times 10^8}{1. \alpha -5.16993\times 10^8},
\end{eqnarray}
where we have used $N=60$. Using typical values of $\lambda$, a parameter $\alpha$ can be fixed by the CMB constraint:
\begin{eqnarray}
\alpha &\sim & -2.5\times 10^{9}\quad{\rm for}\quad\lambda=0.05,
\\
\alpha &\sim & -1.45\times 10^{10}\quad{\rm for}\quad\lambda=0.01 \,.
\end{eqnarray}
Notice that a successful prediction of the density perturbation requires $\alpha$ to be large similar to those found in, e.g.,  Refs.\cite{Calmet:2016fsr,Netto:2015cba}.  The spectral index of curvature perturbation $n_{s}$ and the tensor-to-scalar ratio $r$ are given in terms of the e-foldings $N$:
\begin{eqnarray}
n_{s} &:=& 1-6\epsilon+2\eta \simeq 1-\frac{2}{N}-\frac{9}{2 N^2},
\\
r &:=& 16\epsilon \simeq \frac{12}{N^2} \,.
\end{eqnarray}
Notice that the predictions of our present model are in agreement with those of Ref.\cite{Channuie:2016iyy} where $n_{s}\simeq 0.966$ and $r=0.0033$ for $N=60$.
\section{Dark matter from the HS model}
\label{sec4}
In this section, we solve the renormalization group equations for demonstrating residual effect of the inflation to dark matter in the Higgs-Starobinsky model. In addition, thermal relic abundance of the scalaron from the $R^2$ inflation may has some remnant in the dark matter.

\subsection{Singlet Hermitian Higgs field as candidate of dark matter}
We start with the the gravitation action of the Higgs-Starobinsky model in Jordan frame with the presence of the standard model Lagrangian, $\mathcal{L}_{SM}$.
Introducing auxiliary field $\chi^2 \equiv R$, one  finds,
\begin{eqnarray}
S_J &=& \int d^4x\sqrt{-g}\,\Bigg[ -\frac{1}{2}\,M_P^2\, R - \frac{1}{2}\,\widetilde\xi\,\sigma^2\,R - \frac{1}{2}\,\widetilde\alpha\,\chi^2\,R 
+ \frac{1}{2}\,g^{\mu\nu}\partial_\mu\sigma\,\partial_\nu\sigma
\nonumber\\
&\;& \qquad \qquad \qquad -\, \frac{\widetilde c }{4}\,\chi4 - \frac{\widetilde \kappa}{2}\,\chi^2\,\sigma^2 - \frac{\widetilde\lambda}{4}\,\sigma^4 + \mathcal{L}_{SM} \Bigg],
\label{aux-HS-J}
\end{eqnarray}
where the couplings $\xi$, $\alpha$ and $\lambda$ in Eq. (\ref{HS-J}) are linear combinations of the new coupling constants, $\widetilde \xi$, $\widetilde \alpha$, $\widetilde c$, $\widetilde \kappa$ and $\widetilde \lambda$  as,
\begin{eqnarray}
\xi = \widetilde\xi -\frac{\widetilde \alpha\,\widetilde\kappa}{\widetilde c} \,,\qquad  \alpha = \frac{\widetilde \alpha}{\widetilde c} \,,
\qquad \lambda = \widetilde\lambda - \frac{\widetilde\kappa^2}{\widetilde c} \,.
\label{new-HS-J}
\end{eqnarray}
By varying the action in Eq.(\ref{aux-HS-J}) with respect to the $\chi$ field, this gives $\chi^2 = R$ and we can recover the original action of the Higgs-Starobinsky model in Eq. (\ref{HS-J}). It has been also demonstrated in Ref. \cite{Ema:2017rqn} that there is dual description to choose the choices of the invariant under shift and re-scaling transformation of the auxiliary field $\chi^2$\,. In addition, the $\chi$ field is a scalar mode of the graviton which is called scalaron \cite{Whitt:1984pd,Maeda:1988ab}. It is important to note that the action in Eq.(\ref{aux-HS-J}) automatically generates the same structure of the action as the gauge singlet scalar model of inflation and dark matter \cite{Clark:2009dc,Lerner:2009xg} or Higgs-portal paradigms \cite{Han:2015hda}  as well as the composite NJL model of inflation and dark matter \cite{Channuie:2016iyy}. Moreover, the coupling between scalaron $\chi$ and Higgs $\sigma$ naturally emerges with the choice of re-define new parameters in Eq. (\ref{new-HS-J}). In order to see the behavior of the couplings in different energy scales, the renormalization group analysis is required. This subsection is to study the Higgs-Starobinsky model as candidate of the dark matter. The new scalaron $\chi$ plays the role as dark matter via thermal relic abandance as we will discuss in the next subsection. We impose the energy scale in this study is the elecetroweak scale at the given values of the top qurk mass $m_t= 173$ GeV and vacuum expectation value $v=246$ GeV. The beta functions of the standard model coupling parameters are given in Eqs.(\ref{SM1}), (\ref{SM2}), (\ref{SM3}), (\ref{SM6}) except for Eqs.(\ref{SM4}), (\ref{SM5}) need to modified due to our new tilde parameters $\widetilde\lambda$ and $\widetilde\alpha$\,. The beta functions of the parameters in Eq.(\ref{new-HS-J}) are given by \cite{Clark:2009dc,Lerner:2009xg}
\begin{eqnarray}
\beta_{\widetilde\alpha} &=& \frac{1}{16\pi^2}\,\Bigg[(3 + c_\sigma)\,\widetilde\kappa\left( \widetilde\xi + \frac16\right) + 6\,c_\chi\left( \widetilde\alpha + \frac16\right) \Bigg]\,,
\label{beta-talpha}
\\
\beta_{\widetilde\xi} &=& \frac{1}{16\pi^2}\,\Bigg[\left((6 + c_\sigma)\,\widetilde\lambda + 6\,y_t^2 -\frac32\left(3\,g^2 + (g')^2\right)\right)\left( \widetilde\xi + \frac16\right) + c_\chi\left( \widetilde\alpha + \frac16\right) \Bigg]\,,
\label{beta-txi}
\\
\beta_{\widetilde c} &=& \frac{1}{16\pi^2}\,\Bigg[\frac12\,(c_\sigma^2 +3)\widetilde\kappa + 18\,c_\chi\,\widetilde c^2 \Bigg]\,,
\label{beta-tc}
\\
\beta_{\widetilde\kappa} &=& \frac{1}{16\pi^2}\,\Bigg[4\,c_\sigma\,c_\chi + 6\,(c_\sigma^2 +1)\widetilde c\,\widetilde\kappa \frac32\,(3g^2 + (g')^2)\widetilde\kappa + 6\,y_t^2\,\widetilde\kappa + 6\,c_\sigma\,\widetilde c\widetilde\kappa \Bigg]\,,
\label{beta-tkap}
\\
\beta_{\widetilde\lambda} &=& \frac{1}{16\pi^2}\,\Bigg[18\,(c_\sigma^2 +6)\widetilde\lambda^2 - 6\,y_t^2 +\frac38\left( 2\,g^2 + (g + (g'))^2\right) + \left( 12\,y_t^2 -9\,g^2 - 3\,(g')^2\right)\,\widetilde\lambda +\frac12\,c_\chi^2\,\widetilde\kappa^2 \Bigg]\,,
\label{beta-tlamb}
\end{eqnarray}
where $c_\sigma$ and $c_\chi$ are the suppression factor for the Higgs ($\sigma$) and the scalaron ($\chi$) and we will set them to 1 in the discussion below (see Refs.\cite{Clark:2009dc,Lerner:2009xg,Channuie:2016iyy} for definitions and detail descriptions).
Here we consider the real part of the $\chi$ field only. By solving the beta functions for $\beta_{\widetilde\xi}$ and $\beta_{\widetilde\alpha}$ in Eq.(\ref{beta-txi} ) and Eq.(\ref{beta-talpha}) respectively, it has been demonstrated in Ref. \cite{Channuie:2016iyy} (in the framework of the composite NJL model but it has the same field configurations) that the renormalization group running analysis of the $\widetilde\xi$ and $\widetilde\alpha$ confirms the dominance of the $\chi$ field inflation over the Higgs field with the relation,
\begin{eqnarray}
\widetilde\xi \approx 0.019\,\widetilde \alpha\,.
\end{eqnarray}
Before closing this subsection, we turn to discuss about the electroweak vacuum stability and its perturbative properties of the Higgs-Starobinsky model in the form of action (\ref{new-HS-J}). According to Ref. \cite{Lerner:2009xg}, we find the constraints,
\begin{eqnarray}
\widetilde c >0\,,\qquad \widetilde\lambda >0\,,\qquad \widetilde\kappa >0\,,\quad {\rm or} \quad
 \widetilde\kappa^2 <  \widetilde c \, \widetilde\lambda\,.
\end{eqnarray}
Together with the perturbative conditions up tp Planck scale, one obtains \cite{Lerner:2009xg},
\begin{eqnarray}
\widetilde c < \frac{2\,\pi}{3}\,,\qquad \widetilde\lambda < \frac{2\,\pi}{3}\,,\qquad \widetilde\kappa > 4\,\pi\,.
\end{eqnarray}
These are the constraints of the couplings in Higgs-Starobinsky model as candidate of dark matter. 

\subsection{Relic abundance}
By using the action in Eq.(\ref{aux-HS-J}), it has been shown that the observed mass of dark matter is reproduced by the thermal relic abundance of the $\chi$ field \cite{Clark:2009dc,Lerner:2009xg,Channuie:2016iyy}. We closely follow an analysis and employ the relevant values and results of the parameters for relic abundance in Ref.\cite{Channuie:2016iyy}. The current observed mass of the dark matter is given by \cite{Ade:2013zuv}, 
\begin{eqnarray}
\Omega_{DM}\,h^2 = 0.1199 \pm 0.0027\,,
\end{eqnarray}
where $h \approx 0.7$\,. The crucial parameter for estimating the mass of dark matter is $\widetilde\kappa$ as shown in \cite{Clark:2009dc,Lerner:2009xg,Channuie:2016iyy}. According to the analysis of \cite{Channuie:2016iyy}, to obtain the observed mass of the dark matter (where $M_{DM}$ is represented for mass of dark matter), we find,
\begin{eqnarray}
M_{DM} &=& 61~{\rm GeV}\,,\quad ~{\rm with}\quad \widetilde\kappa = 0.1 \,,
\\
M_{DM} &=& 410~{\rm GeV}\,,\quad {\rm with}\quad \widetilde\kappa = 0.5 \,,
\end{eqnarray}
where the mass formula of dark matter mass is given by \cite{Channuie:2016iyy}
\begin{eqnarray}
M_{DM}^2 = m_\chi^2 +\widetilde\kappa\,v^2/2\,,
\end{eqnarray}
where we have used $m_\chi = m_\sigma/2$ ($m_\sigma$ is Higgs mass) as shown in\cite{Clark:2009dc,Lerner:2009xg,Channuie:2016iyy}. To see the values of the couplings of the original action of the Higgs-Starobinky model, we employ the values of the relevant input parameters in the action (\ref{aux-HS-J}) which have been estimated by \cite{Lerner:2009xg} i.e.,
$\widetilde c =  0.025$\, or \,$\widetilde c =  0.2$\,. Moreover, the solutions of the beta-functions in terms of $\widetilde\xi$, $\widetilde\alpha$ have been done by Ref. \cite{Channuie:2016iyy} and given by
\begin{eqnarray}
\frac{\widetilde\xi}{\widetilde\alpha} = \frac{\widetilde\kappa_M}{18\,\widetilde c_0}\left[ \left( 1-\frac{9\,\widetilde c_0}{8\,\pi^2}\,t\right)^{1/3} - \left( 1-\frac{9\,\widetilde c_0}{8\,\pi^2}\,t\right)\right]\,,
\end{eqnarray}
where $\widetilde\kappa_M$ and $\widetilde c_0$ are average of $\widetilde \kappa$ and initial value of $\widetilde c$\,. We will use above relation to rewrite parameter $\widetilde \xi$ in terms of $\widetilde \alpha$ and we will estimate the values of the $\xi$, $\alpha$ and $\lambda$ parameters by using the definitions in Eq.(\ref{new-HS-J})\,.

\section{conclusion}
In this work, we presented a unified description of inflation and dark matter in the context of the Higgs-Starobinsky (HS) model. The salient feature of this work is to demonstrate that the Higgs-Starobinsky scenario can simultaneously describe inflation and dark matter without introducing new physics beyond standard model. We considered the original action describing the Higgs-Starobinsky model. We started with non-minimal coupling HS action in Jordan frame and transformed it to the Einstein frame using the conformal transformation. We also derived the inflation potential from the gravitational action of non-minimal-Higgs coupling and Starobinsky term in Einstein frame where the $R^2$ term is dominated in the inflationary phase of the universe. For model of inflation, we computed the inflationary parameters and confronted them with Planck 2015 data. We discovered that the predictions of the model are in excellent agreement with the Planck analysis. 

In addition, we considered the Higgs field as a candidate for dark matter. We analyzed the renormalization group equations (RGEs) of Higgs-Starobinsky scenario with the standard model at one-loop level. We made qualitative discussions to the remnants of the quantum effect ($R^2$ term) from Higgs-graviton coupling will be implied by dark matter relic abundance by using the solutions of parameter spaces from RGE analysis.

However, there are some limitations in the present work - for example,  one should complete the RGEs for all scales and solve them
numerically. Moreover, regarding this single
framework, another crucial issue for successful models of inflation is the (pre)reheating mechanism. We plan to investigate this mechanism, within our framework, by following a composite inflationary scenario \cite{Channuie:2016xmq}. In addition, regarding the DM, a thorough analysis of the relic abundance has to be implemented. We hope to address these issues with future investigations.

\section*{Acknowledgements}
DS is supported by Rachadapisek Sompote Fund for Postdoctoral Fellowship, Chulalongkorn University. PC is financially supported by the Institute for the Promotion of Teaching Science and Technology (IPST) under the project of the “Research Fund for DPST Graduate with First Placement”, under Grant No. 033/2557. This work is partially supported by Thailand Center of Excellence in Physics (ThEP).

\end{document}